# Superlattices Consisting of "Lines" of Adsorbed Hydrogen Atom Pairs on Graphene

L. A. Chernozatonskii[a], P. B. Sorokin[a,b], E. E. Belova[a], J. Bruning[c], and A. S. Fedorov[b]

[a]*Emanuel Institute of Biochemical Physics, Russian Academy of Sciences, ul. Kosygina 4, Moscow, 119334 Russia*
*e-mail: cherno@sky.chph.ras.ru*

[b]*Kirensky Institute of Physics, Siberian Division, Russian Academy of Sciences, Akademgorodok, Krasnoyarsk, 660049 Russia*

[c]*Institute of Mathematics, Humboldt University of Berlin, Berlin, 12489 Germany*

The structures and electron properties of new superlattices formed on graphene by adsorbed hydrogen molecules are theoretically described. It has been shown that superlattices of the ($n$, 0) zigzag type with linearly arranged pairs of H atoms have band structures similar to the spectra of ($n$, 0) carbon nanotubes. At the same time, superlattices of the ($n$, $n$) type with a "staircase" of adsorbed pairs of H atoms are substantially metallic with a high density of electronic states at the Fermi level and this property distinguishes their spectra from the spectra of the corresponding ($n$, $n$) nanotubes. The features of the spectra have the Van Hove form, which is characteristic of each individual superlattice. The possibility of using such planar structures with nanometer thickness is discussed.

A monatomic graphite sheet with a hexagonal structure, which is called graphene, has recently attracted strong attention among researchers, because methods for its production on various substrates have been developed [1–4]. It is outstanding due to its unordinary electron properties, because the energy spectrum contains so-called Dirac points, i.e., $K$ points of the connection of band cones to which the Fermi surface degenerates [5]. The carbon nanotubes that are rolled graphene ribbons have either a semimetallic spectrum as in graphene when the allowable wave vector of the ribbon in the direction of its rolling intersects a point $K$ or a semiconducting spectrum in the opposite case. As recently shown [6], vacancy lines on graphene strongly change its spectrum: depending upon the arrangement of these lines, graphene ribbons between them can be considered semiconducting or semimetallic when carriers propagate in the direction perpendicular to the lines, and lines themselves are metallic nanowaveguides.

In this work, we continue to analyze the possibility of forming monatomic superlattices based on graphene, using chemical adsorption of pairs of H atoms on the graphite surface that was recently discovered in experiments on the annealing of physically adsorbed hydrogen molecules [7]. We theoretically show that graphene with "lines" of such pairs is quite energy stable and analyze the band structure of the resulting superlattices. In order to approach real experiment with graphene layers, we take into account a substrate by optimizing the structure of graphene with 2H lines placed on the substrate (graphene layer, according to the Bernal $A$–$B$ scheme, Fig. 1).

Here, we consider only ($n$, 0) and ($n$, $n$) 2H superlattices in which pairs of H atoms are covalently bonded with pairs of neighboring C atoms (see the inset in Fig. 1b), i.e., form one of the stable structures of bonded C–H pairs observed experimentally (cf. Fig. 3 in [7]). In the notation of these superlattices (2HG), the indices ($n$, $m$) used for defining carbon nanotubes [5], superlattices on the graphite sheet with defects [6], and their unit cells are used. It is seen that two $sp^3$ hybridized C atoms with which hydrogen atoms are bonded are ~0.1 nm out of the plane and their bonds with H atoms are slightly different ($\alpha \neq \beta$ in the inset in Fig. 1b). This breaking of the mirror symmetry of pairs of C–H atoms with respect to the $X$ axis occurs due to the Van der Waals interaction of this superlattice with the substrate, which is the graphene layer shifted according to the Bernal $A$–$B$ scheme. It is natural to expect the effect of the breaking of similar symmetry for substrates of other materials with a crystalline structure with the surface different from graphene.

It is worth noting that the effect of such lines of adsorbed H atomic pairs on change in the geometry and electron properties of ($n$, 0) and ($n$, $n$) carbon nanotubes was considered previously [8–11]. Calculation of the periodic 32 C atomic system with one adsorbed hydrogen atom was performed recently [12] in the density functional theory. This calculation shows that, owing to the breaking of graphene symmetry at the Dirac point, an energy gap (1.25 eV) appears and a nondissipative level corresponding to the localization of the electron density near the H atom is formed in this gap neat the Fermi energy.

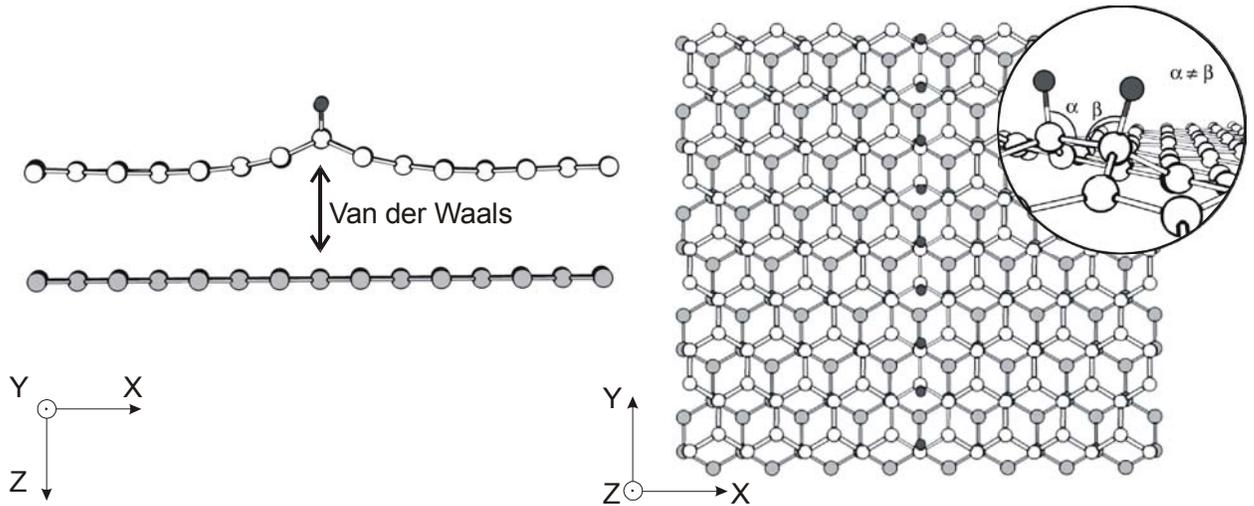

**Fig. 1.** Arrangement of the line of hydrogen atoms covalently bonded with graphene atoms: (a) view in the *Y* axis direction and (b) plan view. Open circles are C atoms of upper graphene on which pairs of H atoms, which are denoted by closed circles and are located in parallel to the *Y* axis, are absorbed. Owing to the molecular interaction of the upper graphene with the substrate, which is the lower graphene layer (C atoms denoted by gray circles), the angles between the bond of the $sp^3$ hybridized C atoms and their bonds with hydrogen atoms are different ($\alpha \neq \beta$, see the inset) due to the breaking of the mirror symmetry with respect to the *X* axis.

All calculations of the electron energy spectra of the 2H graphene Superlattices under the consideration are performed in the framework of the density functional theory with the OpenMX 3.0 program [13, 14], which allows ab initio quantum-mechanical calculations using the local electron density approximation [15-17]. A linear combination of localized pseudoatomic orbitals [14] is used as the basis. A pseudopotential generated be the scheme described in [18] with partial core correction [19] is used as a pseudopotential for carbon and hydrogen. The *s3p3d2* set for carbon and the *s3p3* set for hydrogen are used as valence orbitals. The cutoff radii for the carbon and hydrogen orbitals are taken to be 5.0 and 4.0 au, respectively. A cutoff energy of 150 Ry is taken for numerical integration of the Poisson equation. To obtain the band structure, 100 *k* points in each of the highly-symmetric directions are used. A 50 × 50 × 1 set of *k* points is used to calculate the density of electronic states.

To optimize the structures, the Abel–Tersoff–Brenner molecular dynamics method (parameterization I) [20], which was successfully used in the calculation of carbon nanostructures [21, 22], is used. The molecular dynamics method is used in order to take into account the Van der Waals forces that are determined in this case due to the interaction between neighboring graphene sheets that cannot be taken into account in the framework of the local density approximation of the density functional theory, which is usually used to calculate the electronic characteristics of carbon nanostructures. The molecular interaction potential is taken in the 6–12 standard form [22].

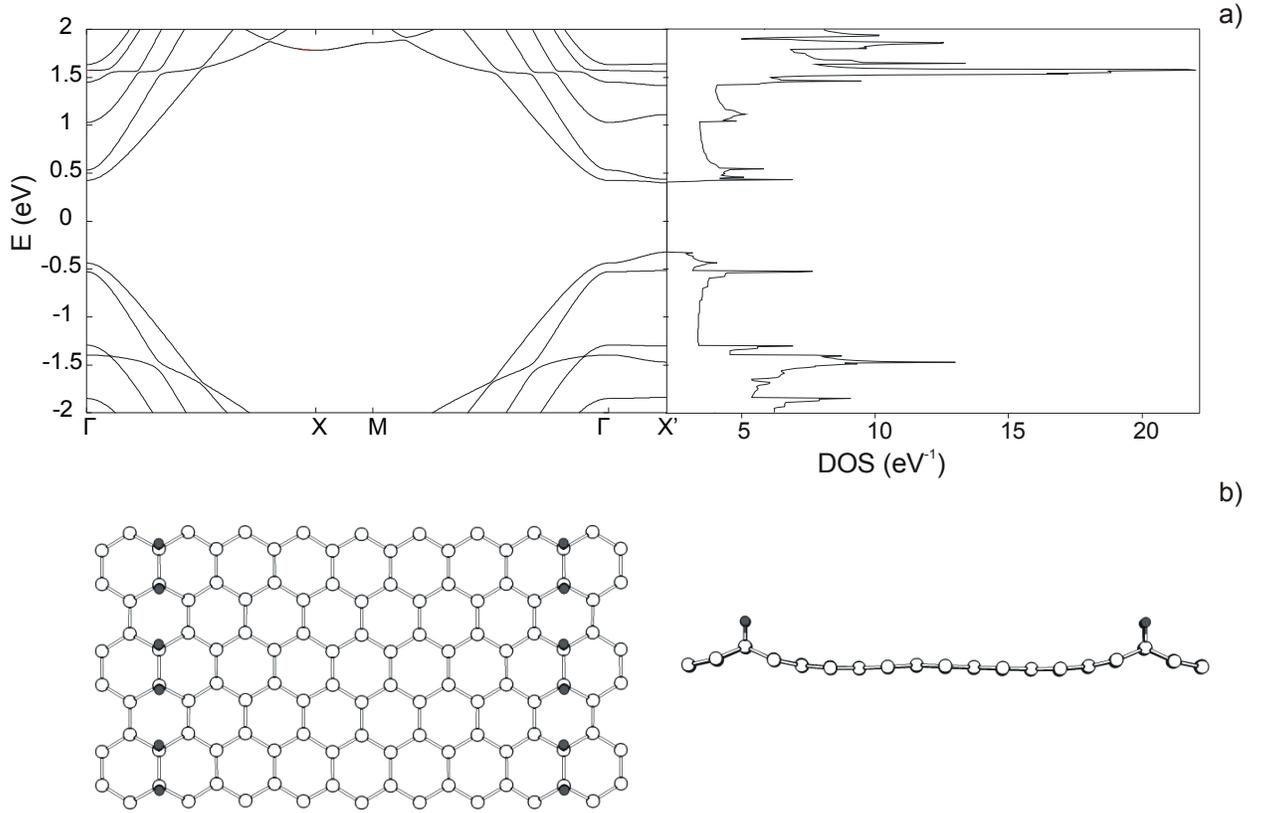

**Fig. 2.** (a) Band spectrum and the density of one-electronic states with Van Hove singularities and (b) the arrangement scheme for 2H lines (plan view and view in the Y axis direction are shown on the left and right, respectively) for the semiconducting 2HG-(7, 0) superlattice with the energy gap $E_g = 0.86$ eV.

**2HG-($n$, 0) "zigzag" superlattices.** First, we consider the systems of the lines of covalently bonded pairs of H atoms of the ($n$, 0) zigzag type. Owing to the breaking of the symmetry of "pure" graphene, they have a semiconducting band structure, i.e., a gap $E_g$ in the spectrum whose values for all superlattices under consideration are presented in the table. According to this table, the gap width depends periodically on the index $n$ (width of the strip between 2H lines) and this dependence is similar to the relation $n = 3q$ for "metallic" nanotubes [5], but the gap does not disappear completely, because a barrier exists on $sp^3$ hybridized C atoms and dissymmetry exists due to the presence of the substrate. The Brillouin zone for the 2HG-($n$, 0) superlattice has a rectangular shape corresponding to its unit cell.

The dispersion curves $E(k)$ and the density of one-electronic states $DOS(E)$ with the Van Hove singularities, which are typical for 2HG-($n$, 0) superlattices, are shown in Fig. 2 for the 2HG-(7, 0) structure. Similar to the (7, 0) nanotube (according to the same density functional theory calculation, its gap is $E_g = 0.321$ eV), the 2HG-(7, 0) system exhibits a direct-band transition in the spectrum at the point $X'$ rather than at the point $\Gamma$. The gap is strongly broadened in the $X$ direction, which indicates the presence of a high energy barrier for the penetration of electrons through the chain of $sp^3$ carbon atoms. The Van Hove peaks exhibit a pronounced energy distribution for each system (cf. Fig. 2a). In this case, there is no simple dependence on the number $n$ for transitions from the valence band to the conduction band $E_{11, 22}$ such as the Kataura plot for carbon nanotubes [23]. We believe that semiconducting 2HG-($n$, 0) superlattices will compete with nanotubes in nanoelectronics and nanooptics, because the production of them is more successful and simpler than the production of nanotubes.

Energy gap width $E_g$ for various indices n of the zigzag superlattice

| $n$ | 5 | 6 | 7 | 8 | 9 | 10 | 11 | 12 | 13 | 14 | 15 |
|---|---|---|---|---|---|---|---|---|---|---|---|
| $E_g$ (eV) | 0.385 | 0.160 | 0.705 | 0.224 | 0.216 | 0.513 | 0.160 | 0.192 | 0.385 | 0.160 | 0.192 |

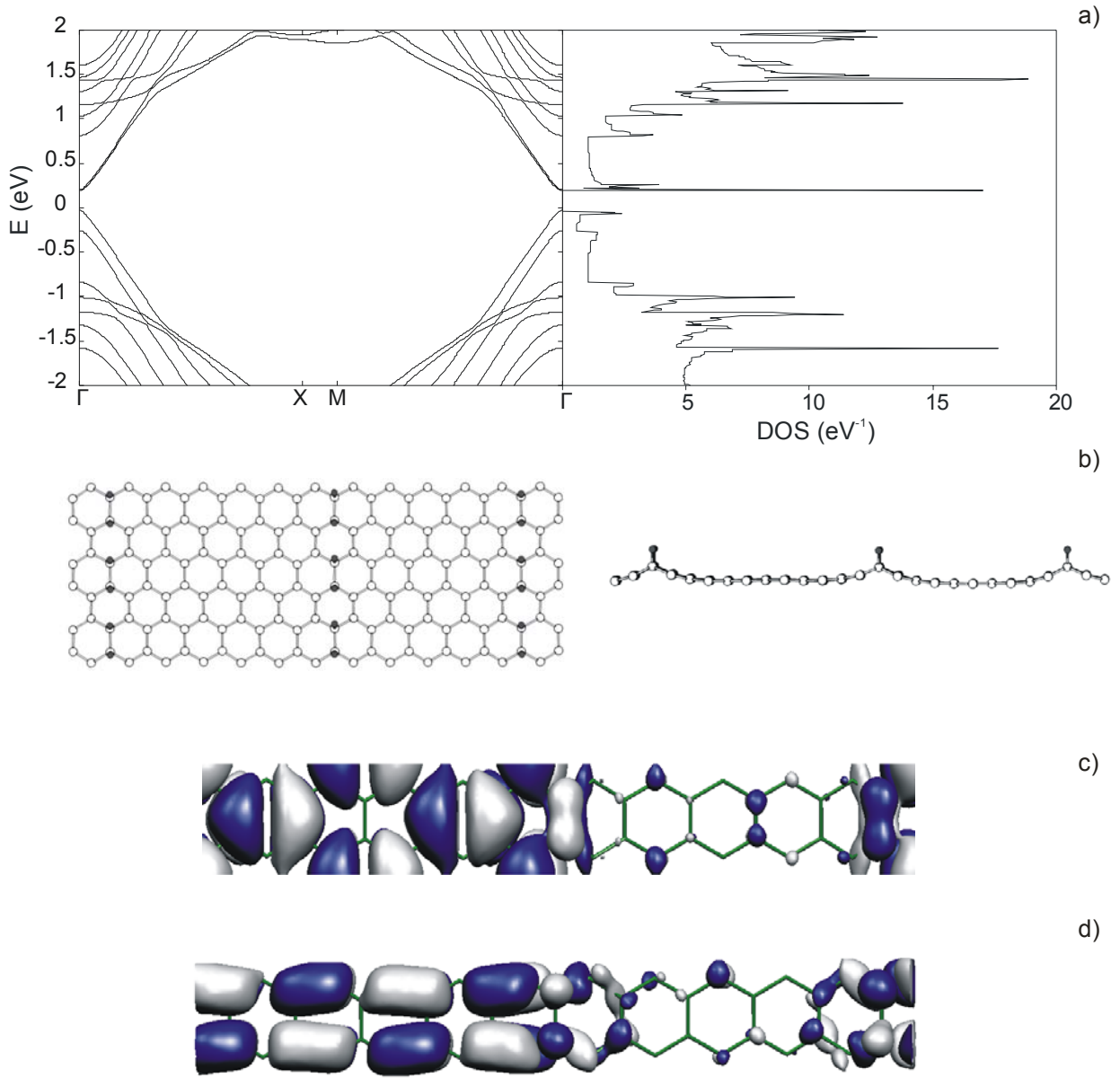

**Fig. 3.** (a) Band spectrum and the corresponding density of one-electronic states with Van Hove singularities, (b) arrangement scheme for 2H lines (plan view and view in the Y axis direction are shown on the left and right, respectively), and (c, d) the distributions of orbitals at the point Γ in a unit cell in the valence band and conduction band, respectively, for the 2HG-(6, 0) + (5, 0) superlattice composed of the (6, 0) + (5, 0) strips. The closed and open points correspond to the molecular orbitals with opposite signs (cutoff value is 0.01).

**2HG-($n$, 0) + ($m$, 0) system.** Since 2HG-($n$, 0) superconducting superlattices with different periods have different energy gaps, it is interesting to investigate more complex systems similar to usual superconducting heterostructures. To this end, the 2HG-(6, 0) + (5, 0) graphene structure with 2H atomic lines alternating at distances of six and five hexagons will be discussed below as an example of the two-period 2HG-($n$, 0) + ($m$, 0) superlattice, see Fig. 3. This choice is explained by the difference between the energy gaps in the corresponding (6, 0) and (5, 0) superlattices, see the table. Figure 3a shows the band spectrum and density of one-electronic states of such two-period superlattice. Similar to one-period 2HG-($n$, 0) systems, the spectrum exhibits the direct band transition at the point Γ with the gap $E_g = 0.232$ eV, which is close to the gap of the 2HG-(6, 0) superlattice. However, in contrast to the corresponding 2HG-(6, 0) and 2HG-(5, 0) characteristics, the density of states at the edge of the conduction band has a higher peak and the number of such Van Hove singularities in the energy scale is larger. Fig. 3 shows the molecular orbitals distributions of (c) the highest occupied molecular orbitals at the point Γ in the valence band and of (d) the lowest unoccupied molecular orbitals at the point Γ in the conduction band. It is seen that electrons are localized near the 2H (6, 0) strip, which, being rolled, forms the metallic carbon (6, 0) nanotube [5]. The more "dielectric" (5, 0) strip is almost empty.

The consideration of this effect shows that an electron waveguide, a quantum nanometer wire with a thickness of "one atom," can be created by enclosing a "quasi-metallic" strip [e.g., the 2H-(6, 0) strip] on graphene with "dielectric" strips [e.g., the 2H-(5, 0) strips] from two sides.

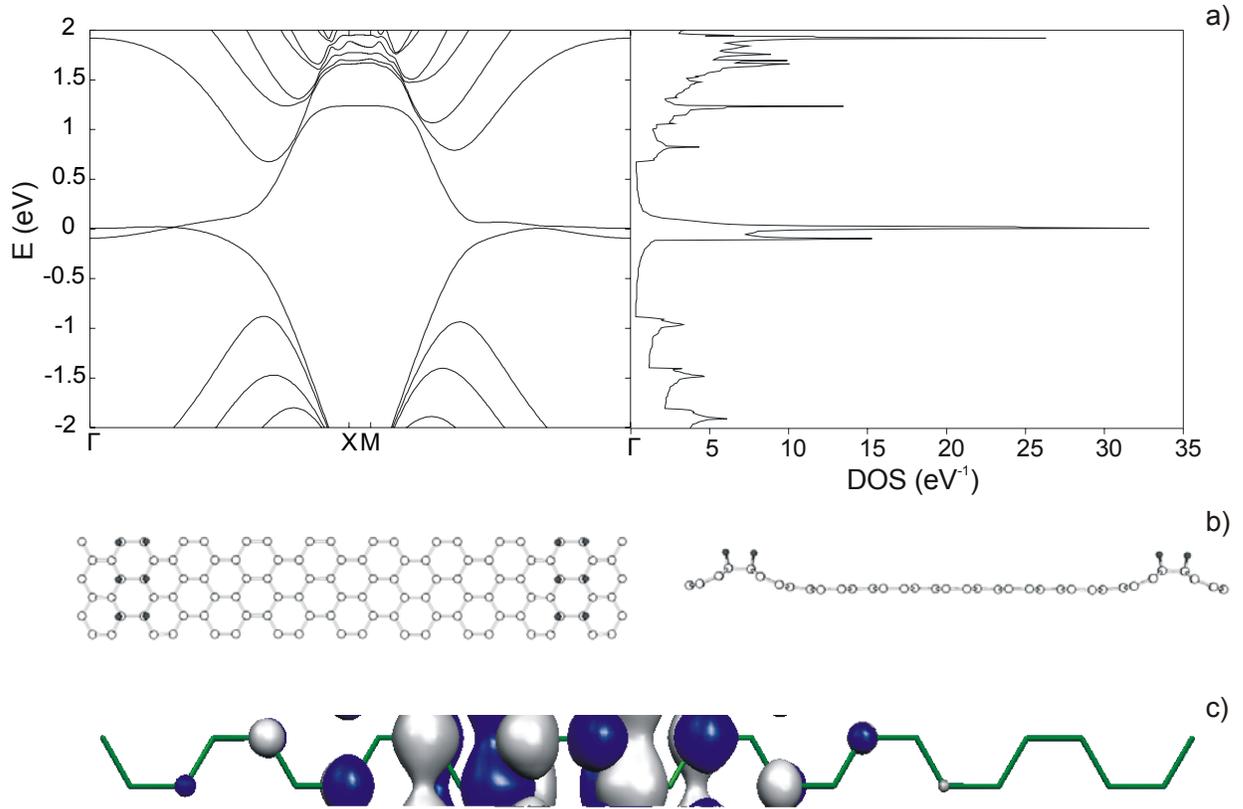

**Fig. 4.** (a) Band spectrum and the corresponding density of one-electronic states with Van Hove singularities, (b) arrangement scheme for 2H lines (plan view and view in the Y axis direction are shown on the left and right, respectively), and (c) the distributions of electron molecular orbitals at the Fermi energy at the point Γ (cutoff value is 0.025) for the 2HG-(7, 7) superlattice.

**2HG-(n, n) armchair superlattices.** We now study the graphene structures with 2H atomic lines forming periodically arranged (n, n) armchair strips: pairs of H atoms are located on neighboring C atoms, forming "staircase stairs," see Fig. 4. The band structures of such superlattices are similar to each other. Let us consider their form in more detail for the 2HG-(7, 7) superlattice, see Fig. 4b.

In the Y direction (ΓX region) parallel to the C–H staircases, the character of the spectrum is similar to the band structure of a graphene ribbon [24] or (n, n) armchair nanotubes [5]. However, the point of the intersection of branches is shifted towards the middle of the ΓX segment and their flat parts provide the high density of states at the Fermi level. Owing to this circumstance, the (n, n) strips in the superlattices under consideration are significantly more metallic as compared to the electronic characteristics of the corresponding armchair nanotubes. Here, the molecular orbitals of electrons at $E_F$ is localized near the C–H staircase (see Fig. 4c) and this localization is similar to its localization on the dislocation line in graphene [6] or at the edges of the graphene ribbon [24].

In the X direction (XM region) perpendicular to the C–H staircases, a wide dielectric gap appears in the spectrum. For this reason, a high energy barrier created by pairs of $sp^3$ hybridized C atoms prevents the motion of the current carriers along the X direction.

Thus, it has been shown in this work that covalently bonded pairs of hydrogen atoms on graphene in the form of the superlattice strips with nanometer periods significantly change the semimetallic spectrum of pure graphene: 2HG-(n, 0) superlattices are semiconductors with a gap depending on their period, whereas 2HG-($n$, $n$) superlattices are substantially metallic (with dielectric nano-partitions).

The Van Hove singularities in the spectra take the form characteristic of each separate superlattice. For this reason, Raman scattering methods allow one to identify and control each structure, which must have spectral "fingerprint" characteristic only of it in the Raman spectra of carbon nanotubes [5] and pure graphene layers [25]. Van Hove peaks in the band structure of these superlattices must be manifested in all features of nonlinear optics (luminescence, fluorescence, etc. [5]) that are characteristic of carbon nanotubes.

We have also shown that the electron density near the Fermi energy is localized near the 2H lines. Hence, electronic nanowaveguides and nanoheterostructures and, therefore, nanoelectronic devices based on them, can be obtained by creating lines of pairs of H atoms adsorbed on graphene, superlattices with lines of pairs of atoms of other elements can obviously be created on graphene. In particular, the fluoridation of 2F vapors along the lines must also lead to structures with similar electron characteristics. However, the use of $H_2$ molecules seems to be the most appropriate for implementation according to the experiment reported in [7]. The nanoprint method will allow the production of quantum electronic chips on a single graphite sheet, which are analogues of the chips of integrated optics (electron waveguide–optical waveguide).

We are grateful to the Joint Supercomputer Center of the Russian Academy of Sciences for the possibility of using a cluster computer for quantum-chemical calculations, to I.V. Stankevich and L. Biro for stimulating discussions. The electron density was visualized using the Molekel 4.0 program. This work was supported by the Russian Foundation for Basic Research (project no. 05-02-17443) and Deutsche Forschungsgemein-schaft/Russian Academy of Sciences (DFG/RAS, project no. 436 RUS 113/785).

## REFERENCES

1. C. Oshima and A. Nagashima. J. Phys.: Condens. Matter **9**, 1 (1997).
2. K.S. Novoselov, A.K. Geim, S.V. Morozov *et al.*, Science **306**, 666 (2004).
3. C. Berger, Z. Song, T. Li *et al.,* J. Phys. Chem. [Sect. B] **108**, 19912 (2004).
4. K.S. Novoselov, A.K. Geim, S.V. Morozov *et al.*, Nature **438**, 198 (2006)
5. *Carbon Nanotubes: Synthesis, Structure, Properties,and Applications*, Ed. by M. S. Dresselhaus, G. Dresselhaus,and Ph. Avouris (Springer, Berlin, 2001), Topics in Applied Physics, Vol. **80**.
6. L. A. Chernozatonskii, P. B. Sorokin, E. E. Belova, et al., Pis'ma Zh. Eksp. Teor. Fiz. **84**, 141 (2006) [JETP Lett. **84**, 115 (2006)], cond-mat/0611334
7. L. Hornekær, Ž. Šljivančanin, W. Xu, R. Otero *et al.,* Phys. Rev. Lett. **96**, 156104 (2006)
8. D .Srivastava, D.W. Brenner, J.D. Schall *et al.,* J. Phys. Chem. [Sect. B] **103**, 4330 (1999)
9. I. V. Zaporotskova, A. O. Litinskiœ, and L. A. Chernozatonskiœ, Pis'ma Zh. Eksp. Teor. Fiz. **66**, 799 (1997) [JETP Lett. **66**, 841 (1997)]
10. E. Gal'pern, I. Stankevich, A. Chistyakov, and L. A. Chernozatonskiœ, Izv. Ross. Akad. Nauk, Ser. Khim. **11**, 2061 (1999).
11. O. Gülseren, T. Yildirim, and S. Ciraci, Phys. Rev. Lett. **87**, 116802 (2001)
12. E.J. Duplock, M. Scheffler, P.J.D. Lindan, Phys. Rev. Lett. **92**, 225502 (2004).
13. T. Ozaki, Phys. Rev. **B 67**, 155108 (2003).
14. T. Ozaki and H. Kino, Phys. Rev. **B 69**, 195113 (2004).
15. W. Kohn and L. J. Sham, Phys. Rev. [Sect. A] **140**, 1133 (1965).
16. P. Hohenberg and W. Kohn, Phys. Rev. [Sect. B] **136**, 864 (1964).
17. D. M. Ceperley and B. J. Alder, Phys. Rev. Lett. **45**, 566 (1980).
18. N. Troullier and J.L. Martins, Phys. Rev. **B 43**, 1993 (1991).
19. S.G. Louie, S. Froyen, and M.L. Cohen, Phys. Rev. **B 26**, 1738 (1982).
20. D D.W. Brenner, Phys. Rev. **B 42**, 9458 (1990).
21. A. Garg, S.B. Sinnott, Phys. Rev. **B 60**, 786 (1999).
22. S.B. Sinnott, O.A. Shenderova, C.T. White and D.W. Brenner, Carbon **36**, 1 (1998).
23. H. Kataura, Y. Kumazawa, Y. Maniwa *et al.,* Synthetic Metals **103**, 2555 (1999).
24. Y.-W. Son, M. L. Cohen , and S. G. Louie. Nature **444**, 347 (2006).
25. C. Ferrari, J.C. Meyer, V. Scardaci *et al.,* Phys. Rev. Lett. **97,** 187401 (2006).